\documentclass[12pt,preprint]{aastex}

\newcommand{\hst}{\textit{HST\ }}

\begin{document}
\title{New Constraints on Additional Satellites of the Pluto System}

\author{A.~J.~Steffl\altaffilmark{1}, M.~J.~Mutchler\altaffilmark{2},
  H.~A.~Weaver\altaffilmark{3}, S.~A.~Stern\altaffilmark{4},
  D.~D.~Durda\altaffilmark{1}, D.~Terrell\altaffilmark{1},
  W.~J.~Merline\altaffilmark{1}, L.~A.~Young\altaffilmark{1},
  E.~F.~Young\altaffilmark{1}, M.~W.~Buie\altaffilmark{5}, and
  J.~R.~Spencer\altaffilmark{1}}

\email{steffl@boulder.swri.edu}

\altaffiltext{1}{Southwest Research Institute, Department of Space
  Studies, Boulder, CO 80302}

\altaffiltext{2}{Space Telescope Science Institute, 3700 San Martin
  Drive, Baltimore, MD 21218.}

\altaffiltext{3}{The Johns Hopkins University Applied Physics
  Laboratory, Space Department, 11100 Johns Hopkins Road, Laurel, MD
  20723-6099.}

\altaffiltext{4}{Southwest Research Institute, Space Science and
  Engineering Division, 1050 Walnut Street, Suite 400, Boulder, CO
  80302.}

\altaffiltext{5}{Lowell Observatory, 1400 W. Mars Hill Road,
  Flagstaff, AZ 86001.}

\shorttitle{Constraints on Additional Plutonian Satellites}
\shortauthors{Steffl et al.}
\slugcomment{Revised submission to the Astronomical Journal}

\begin{abstract}
  Observations of Pluto and its solar-tidal stability zone were made
  using the Advanced Camera for Surveys' (ACS) Wide Field Channel
  (WFC) on the {\it Hubble Space Telescope} on UT 2005 May 15 and UT
  2005 May 18. Two small satellites of Pluto, provisionally designated
  S/2005~P~1 and S/2005~P~2, were discovered, as discussed by
  \cite{Weaveretal06} and \cite{Sternetal06}.  Confirming observations
  of the newly discovered moons were obtained using the ACS in the
  High Resolution Channel (HRC) mode on 2006 Feb 15
  \citep{Mutchleretal06IAUC}. Both sets of observations provide strong
  constraints on the existence of any additional satellites in the
  Pluto system. Based on the May 2005 observations using the ACS/WFC,
  we place a 90\%-confidence lower limit of $m_{_V}=26.8$
  ($m_{_V}=27.4$ for a 50\%-confidence lower limit) on the magnitude
  of undiscovered satellites greater than 5\arcsec\ ($1.1 \times
  10^5$~km) from Pluto. Using the 2005 Feb 15 ACS/HRC observations we
  place 90\%-confidence lower limits on the apparent magnitude of any
  additional satellites of $m_{_V}=26.4$ between 3\arcsec--5\arcsec\ 
  ($6.9 \times 10^4$--$1.1 \times 10^5$~km) from Pluto, $m_{_V}=25.7$
  between 1\arcsec--3\arcsec\ ($2.3 \times 10^4$--$6.9 \times
  10^4$~km) from Pluto, and $m_{_V}=24.$ between 0\farcs3--1\arcsec\ 
  ($6.9 \times 10^3$--$2.3 \times 10^4$~km) from Pluto. The
  90\%-confidence magnitude limits translate into upper limits on the
  diameters of undiscovered satellites of 29~km outside of 5\arcsec\ 
  from Pluto, 36~km between 3\arcsec--5\arcsec\ from Pluto, 49~km
  between 1\arcsec--3\arcsec\ from Pluto, and 115~km between
  0\farcs3--1\arcsec\ for a comet-like albedo of $p_{_V}=0.04$. If
  potential satellites are assumed to have a Charon-like albedo of
  $p_{_V}=0.38$, the diameter limits are 9~km, 12~km, 16~km, and
  37~km, respectively.
\end{abstract}



\section{Introduction}

Since its discovery in 1930 by Tombaugh \citep{Slipher30}, there have
been surprisingly few published searches for satellites of Pluto. The
first search was made at Lowell Observatory in February-March 1930,
immediately following Pluto's discovery \citep{Tombaugh60}. It failed
to discover Charon or any other satellite. Kuiper and Humason, working
independently, conducted satellite searches in January 1950
\citep{Kuiper61}. Using photographic plates from the two searches,
Kuiper established magnitude limits of Plutonian satellites of
$m_{_P}=19$ between 0\farcs3--2\arcsec\ from Pluto and $m_{_P}=22.4$
for the region from 2\arcsec\ from Pluto to the edge of the stability
zone. Curiously, despite the fact that the V magnitude of Charon in
1950 was about 17.5 \citep{Sternetal91} and that it was near its
maximum northern elongation of 0\farcs8 from Pluto at the time of
Kuiper's observations \citep{Reaves97}, no satellites were detected,
although the presence of unresolved Charon in Kuiper's data may have
resulted in his anomolously large measurement of Pluto's diameter
\citep{Marcialis:merline98}. The first 50 years of Pluto-Charon
observations are reviewd by \cite{Marcialis97}.

More recently, \cite{Sternetal91} searched for satellites of Pluto out
to the edge of Pluto's solar-tidal stability region using the
Michigan-Dartmouth-MIT Observatory at Kitt Peak, Arizona and the 2.1m
Struve telescope at McDonald Observatory in Texas. Using a
non-standard filter passband consisting of two separate transmission
peaks at 5000\AA\ (FWHM = 350\AA) and 6575\AA\ (FWHM = 225\AA), they
placed 90\%-confidence limits on satellites brighter than
$m=20.6\pm0.5$ from 6\arcsec--10\arcsec\ from Pluto and $m=22.6\pm0.5$
for angular separations greater than 10\arcsec\ from Pluto. These
90\%-confidence limiting magnitudes were improved by
\cite{Sternetal94} to $m_{_V}=21.7$ between 1\arcsec--2\arcsec\ from
Pluto and $m_{_V}=21.9$ between 2\arcsec--10\arcsec\ from Pluto
through analysis of archival images from the {\it Hubble Space
  Telescope}. No satellites (other than Charon) were detected in
either of these searches. \cite{Nicholson:gladman06} have also
reported results from a Plutonian satellite search using the Hale 5-m
telescope in June of 1999. They searched Pluto's entire Hill sphere
and placed a 50\%-confidence detection limit of $m_{_R}=25.0\pm0.2$ on
additional satellites appearing more than $\sim$4\arcsec\ from Pluto,
the point at which scattered light significantly degrades the
sensitivity of their search. For potential undiscovered satellites
with a solar \vr\ color, this limit translates into
$m_{_V}=25.4\pm0.2$.

Previous satellite searches have used both the Hill radius, $r_{_H}$,
and the stability radius, $r_{_S}$, to define the outer edge of the
search region.  Both of these radii derive from analytic solutions to
the restricted three-body problem (i.e., a massless particle moving in
the gravitational influence of the Sun with mass, $M_{\sun}$, and a
planet with mass, $M_p$). For a satellite to be gravitationally bound
to a planet, it must have sufficiently low energy, so that its
zero-velocity surface is closed. The largest, closed zero-velocity
surface is the Hill sphere, whose radius is given by:

\begin{equation}
  \label{hill_eqn}
  r_{_H} = a_p(1-e_p)\left(\frac{M_p}{3M_{\sun}}\right)^{1/3}
\end{equation}

\noindent where $a_p$ is the semi-major axis of the planet's orbit and
$e_p$ is the planet's orbital eccentricity \citep{Hamilton:burns92}.
Somewhat less well-known is the solar-tidal stability radius given by
Szebehely's stability criterion: $r_{_S}$=1/3 $r_{_H}$
\citep{Szebehely67,Szebehely78}. Satellite orbits with semi-major
axis, $a_s$, less than $r_{_S}$ will be stable over long timespans,
whereas for orbits with $r_{_S} < a_s < r_{_H}$, instability can
develop.  (At distances greater than $r_{_H}$, the satellite is no
longer bound gravitationally to the planet.)  Analytical arguments by
\cite{Hamilton:krivov97} showed that satellites on initially-circular
orbits will become unstable if $a_s \gtrsim 0.53 \, r_{_H}$ for
prograde orbits or $a_s \gtrsim 0.69\, r_{_H}$ for retrograde orbits.
Numerical simulations also show that the orbits of satellites with
$a_s \gtrsim 0.4\,r_{_H}$ for prograde orbits or $a_s \gtrsim
0.7\,r_{_H}$ for retrograde orbits become unstable on timescales of
$\sim10^6$ years \citep{Carubaetal02,Nesvornyetal03}. Thus,
Szebehely's radius, $r_{_S}$, provides a better approximation to the
size of the satellite orbital stability zone than the Hill radius,
$r_{_H}$, and, following \cite{Sternetal91}, we adopt $r_{_S}$ when
referring to the region of stable orbits in the discussion below. For
Pluto, $a_p=39.5$ astronomical units (AU), $e_p=0.248$, and the
combined mass of the Pluto-Charon system, $M_p$, is $1.4570 \pm 0.0009
\times 10^{22}$ kg \citep{Buieetal06}, yielding a stability radius of
$r_{_S}=2.0\times 10^6$~km.

Pluto's first discovered moon, Charon \citep{Christy:harrington78},
orbits with a semi-major axis, $a_0$, of $19,571 \pm 4$ km
\citep{Buieetal06}, within the inner 1\% of Pluto's orbital stability
region \citep{Sternetal91}.  Dynamical interactions with Charon cause
satellite orbits between 0.47~$a_0$ and $\sim$2.0~$a_0$ to become
unstable \citep{Sternetal94}.  Thus, satellites may be found orbiting
Pluto out to a distance of 0.47~$a_0$ and orbiting the Pluto-Charon
barycenter between $\sim$2.0~$a_0$ and the stability radius, $r_{_S}$,
although satellites on orbits closer to Pluto than Charon would be
difficult to explain unless they post-date Charon's outward orbital
migration. The relatively large size of this region, the relatively
bright limits reached by previous searches (except for
\cite{Nicholson:gladman06}, which was not published when we submitted
our proposal), and the upcoming launch of the {\it New Horizons} Pluto
mission in January 2006 motivated our search for additional satellites
of Pluto.

\section{Observations}

\begin{figure}
  \plotone{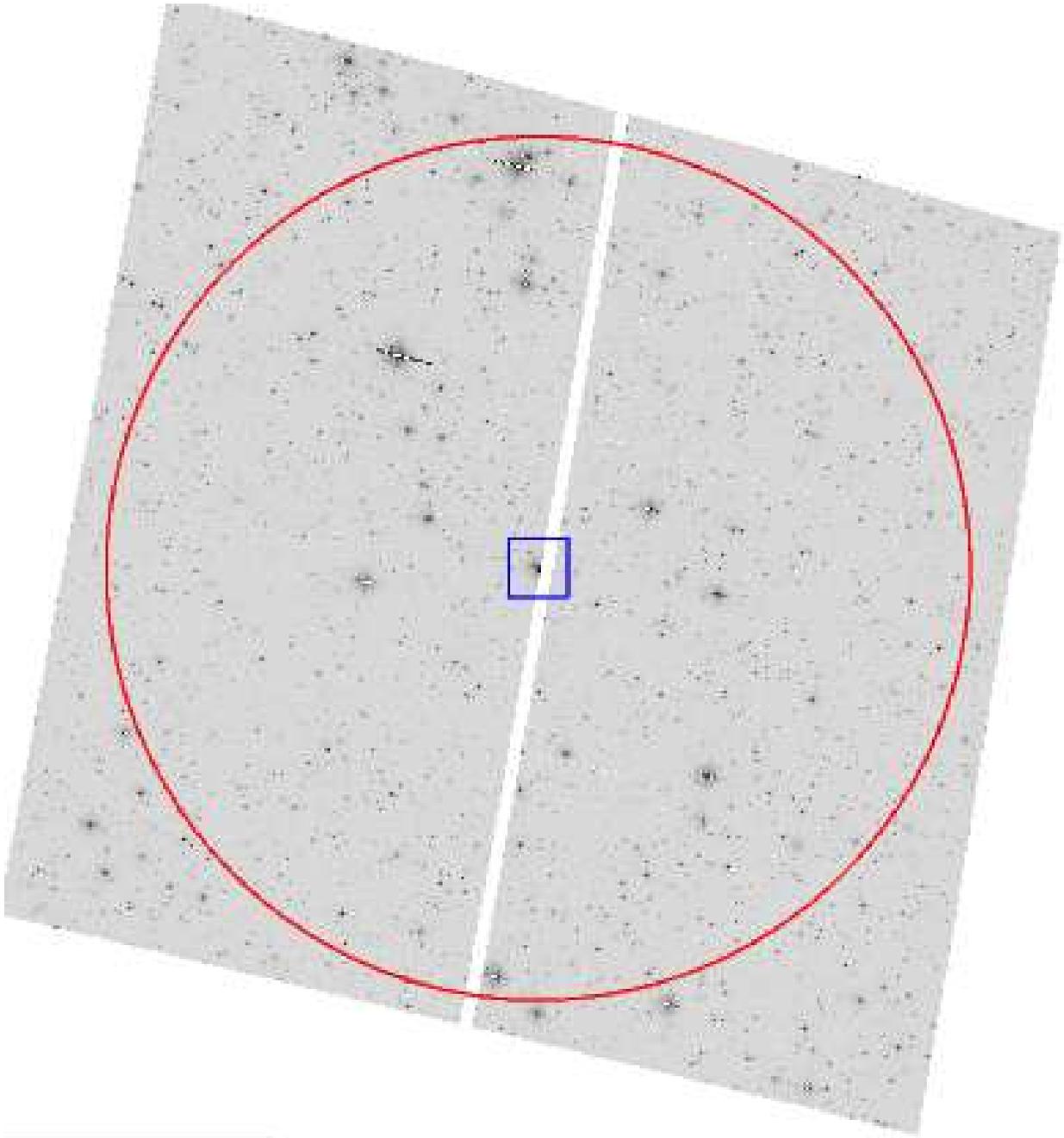}
  \caption[]{ACS/WFC image of Pluto and its stability region on UT 2005 
    May 15. The image has been corrected for the geometric distortion
    of the ACS and rotated so that north is up and east to the left.
    The ACS/WFC consists of two independent $4096 \times 2048$-pixel
    CCDs butted together to form an effective $4096 \times 4096$-pixel
    CCD with an approximately 50-pixel gap separating the two chips.
    The large field of view of the ACS/WFC ($202\arcsec \times
    202\arcsec$) allowed Pluto's entire stability zone, delineated by
    the large circle, to be imaged in a single exposure. The square at
    the center of the image represents the approximate size and
    location of the image in Fig.~\ref{hrc_image}.
    \label{wfc_image}}
\end{figure}

We conducted a search for additional satellites of Pluto using the
F606W filter (``broad V'') in the Wide Field Channel (WFC) of the
Advanced Camera for Surveys (ACS) on the {\it Hubble Space Telescope}
(\hst) during two separate visits on UT 2005 May 15 and UT 2005 May 18
(Guest Observer program 10427). Observational details can be found in
Table~\ref{obs_table}. The ACS/WFC has a field of view of $202\arcsec
\times 202\arcsec$ with a plate scale of 0\farcs049 pixel$^{-1}$. This
is well-matched to the angular size of Pluto's stability region
(185\arcsec\ in angular diameter, as seen from Earth in May 2005),
allowing the entire stability region to be imaged in a single \hst
pointing. A total of five images were obtained per \hst visit: one
short (0.5 sec) exposure with Pluto and Charon unsaturated and four
long (475 sec) exposures. Pluto's apparent motion, seen from \hst and
averaged over the visibility period, is 4\farcs2 h$^{-1}$. This caused
stars to appear as streaks $\sim$11 pixels long, while unresolved
objects moving with Pluto appeared as point sources in all four long
images. A sample long exposure from this dataset can be seen in
Fig.~\ref{wfc_image}.  \placetable{obs_table} \placefigure{wfc_image}

\begin{figure}
  \plotone{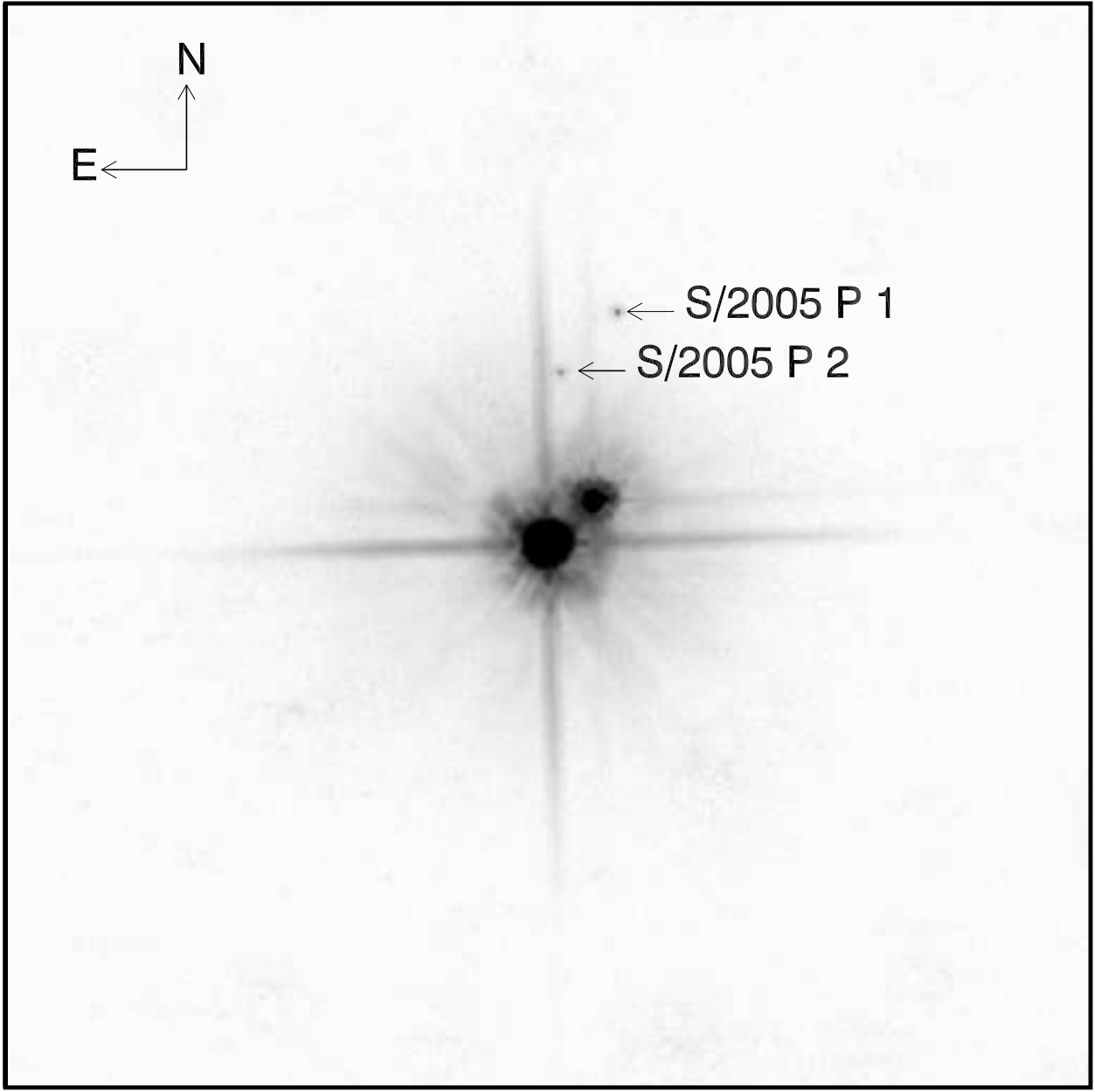}
  \caption[]{Drizzle combined ACS/HRC image of the Pluto system on 
    2006 February 15. This image shows a 512x512 pixel region near the
    center of the ACS/HRC detector. Pluto is centered on the image,
    while Charon is located 0\farcs77 from Pluto at a position angle
    of 313$^{\circ}$. P1 (2\farcs86 from Pluto at a position angle of
    343$^{\circ}$) and P2 (2\farcs03 from Pluto at a position angle of
    356$^{\circ}$) can be clearly seen. No other satellites are
    detected.
    \label{hrc_image}}
\end{figure}

With the detection of two additional satellites of Pluto in the
ACS/WFC observations \citep{Weaveretal06}, we obtained two additional
\hst visits to the Pluto system on 2006 February 15 and 2006 March 2
from the Director's Discretionary Time (GO/DD program 10774). These
observations were designed to confirm the existance of the two
satellites \citep{Mutchleretal06IAUC} and were obtained using the ACS
High Resolution Channel (HRC). The observations in February 2006
consisted of four 475-second integrations taken with the F606W filter
with additional 1-second integrations (to allow for accurate image
registration with Pluto and Charon unsaturated) at each position in a
four point dither pattern. The final drizzle-combined image from the
February 15, 2006 observations, with labels indicating the positions
of S/2005~P~1 and S/2005~P~2, is shown in Fig.~\ref{hrc_image}.
\placefigure{hrc_image}

The observations in March 2006 were also designed to measure the \bv\ 
colors of the satellites, and therefore alternated between 145-second
integrations using the F606W filter and 475-second integrations using
the F435W (Johnson B) filter. As such, the March 2 observations are
less sensitive to faint satellites and were not used in the subsequent
analysis. Some observational details about the ACS/HRC observations
are also presented in Table~\ref{obs_table}.

\section{Data Analysis}

To estimate the sensitivity of our satellite search using the May 2005
ACS/WFC data, we generated synthetic point-spread functions (PSFs) at
400 locations in the plane of the sky using the Tiny Tim v6.3 software
package \citep{Krist:hook04}, with the assumption that the sources
have the same spectral distribution as the Sun. These PSFs were
randomly-spaced in separation and position angle from Pluto and were
scaled to uniformly span the range in WFC STMAG magnitudes of
25.5--29.5. To ensure proper sub-pixel alignment of the synthetic PSFs
in the geometrically-distorted ACS images (i.e. FLT images) the PSFs
were subsampled by a factor of five, resampled at the proper pixel
locations, rebinned to normal size, and then convolved with the CCD
charge diffusion kernel generated by the Tiny Tim program. Independent
Poisson noise was applied to the synthetic PSFs before they were
added, at the appropriate locations, to each of the four deep
(475-second) exposures in the two ACS/WFC visits. The four images were
then ``drizzled'' together using the ``Multidrizzle'' procedure
supplied with the PyRAF software package \citep{Koekemoeretal02}. In
the drizzle procedure, the individual images are corrected for the
geometric distortion of the ACS instrument, rotated so that north is
up and east to the left, sky background subtracted, co-registered
relative to Pluto, and combined using a median filter. In addition,
pixels that have anomalously low sensitivity, that have high dark
counts, or are saturated are excluded. The median combination removes
artifacts, such as cosmic ray events or star trails, that do not
appear in same position on the plane of the sky in at least two of the
images. We then visually searched the final drizzle-combined image for
objects (real or synthetic) with a PSF-like appearance.

Adding synthetic PSFs to the data before conducting the actual
satellite search results in a more accurate estimate of the limiting
magnitude of the search (since the conditions during the satellite
search and the limiting magnitude estimation are identical). However,
there is a small chance that one of the synthetic PSFs will be
coincident with a real source, thus preventing the detection of the
real source. Given the size of the ACS/WFC images and the relatively
compact nature of the ACS/WFC PSF ($0.796\pm0.003$ of the total flux
from a point source is contained within a circle of radius 3 pixels
for the ACS/WFC using the F606W filter \cite{Siriannietal05}), only
0.06\% of the pixels in the ACS/WFC images change by more than 0.1
times the standard deviation in the sky background when the synthetic
PSFs are added.  Thus, we feel the benefit of obtaining a more
accurate estimate of the limiting magnitude vastly outweighs the small
risk of masking a real object with a synthetic PSF.

Visual identification of point sources is generally more reliable than
identification by automated detection algorithms. However, it can also
be more subjective and prone to operator error/fatigue. To minimize
the possibility that a {\it bona fide} satellite of Pluto would be
missed on account of operator error or some systematic error
introduced via the drizzle analysis procedure, the data were searched
a second time using an independent technique: the four deep exposures
were manually co-registered, displayed to the screen, and then cycled
rapidly between images at roughly 15 frames per second. As a result,
stars would appear as trails moving through the displayed region and
cosmic ray events and bad pixels would appear and disappear. Objects
co-moving with Pluto (whether real satellites or synthetic PSFs) would
appear in the same location in each of the four frames. Although this
technique proved to be somewhat less sensitive than the drizzle
combination technique, it yielded consistent results.

The field of view of the ACS/HRC is $29\arcsec \times 26\arcsec$,
compared to $202\arcsec \times 202\arcsec$ with the ACS/WFC. With the
smaller field of view, there is an increased risk that a synthetic PSF
added to the data will be coincident with a real source in the data,
thus preventing the detection of the real source. To avoid this
possibility, data from the February 2006 ACS/HRC observations were
first searched for satellites using the drizzle technique without the
addition of synthetic PSFs. The data were then analyzed a second time,
this time with synthetic PSFs added to provide an estimate of the
sensitivity. A total of 200 PSFs, uniformly spaced between magnitudes
24.5 and 28.5, were added at random locations in each of two annuli
centered on Pluto: one extending from 1\arcsec--3\arcsec\ and the
other from 3\arcsec--5\arcsec. Since placing so many synthetic PSFs in
such a small area would result in a high probability of overlapping
PSFs, the ACS/HRC data was analyzed 10 separate times, with only 20
randomly-selected PSFs in each annulus for each analysis run. After the
synthetic PSFs were added, the ACS/HRC images were drizzle-combined
and the resulting image was visually inspected for point sources in a
manner similar to the ACS/WFC images.

Finally, to estimate the limiting magnitude within 1\arcsec\ of Pluto,
we placed 12 synthetic PSFs in a ring at an angular distance of
0\farcs5 from Pluto, using the techniques described above. The
magnitude of the individual PSFs in the ring pattern was then varied
until at least one of the PSFs could no longer be easily identified.
Although this technique is not as statistically rigorous, it provides
a reasonable estimate of the limiting magnitude in this region.
Scattered light from Pluto prevents us from assigning meaningful upper
limits within $\sim$0\farcs3 of Pluto.

\section{Results and Discussion}

As mentioned above, two satellites of Pluto, provisionally designated
S/2005~P~1 and S/2005~P~2 (hereafter P1 and P2), were discovered
during the analysis of the May 2005 ACS/WFC images
\citep{Weaveretal06}. During the discovery epoch, P1 had an apparent
magnitude of $m_{_V}=22.93\pm0.12$ and P2 had an apparent magnitude of
$m_{_V}=23.38\pm0.17$ \citep{Weaveretal06}. Analysis of the discovery
observations and archival \hst observations yielded provisional orbits
with semi-major axes of $64,780\pm88$~km for P1 and $48,675\pm121$~km
for P2 \citep{Buieetal06}. The orbits of P1 and P2 are circular (or
nearly so) and co-planar with Pluto's other large moon, Charon,
implying these moons share a giant impact origin \citep{Sternetal06}.
This hypothesis is supported by the observation that P1 and P2 are
essentially neutral in color with \bv\ values of $0.653\pm0.026$ for P
1 and $0.654\pm0.065$ for P2 \citep{Sternetal06IAUC}. No other
satellites were detected, out to the edge of Pluto's stability region.

\begin{figure}
  \plotone{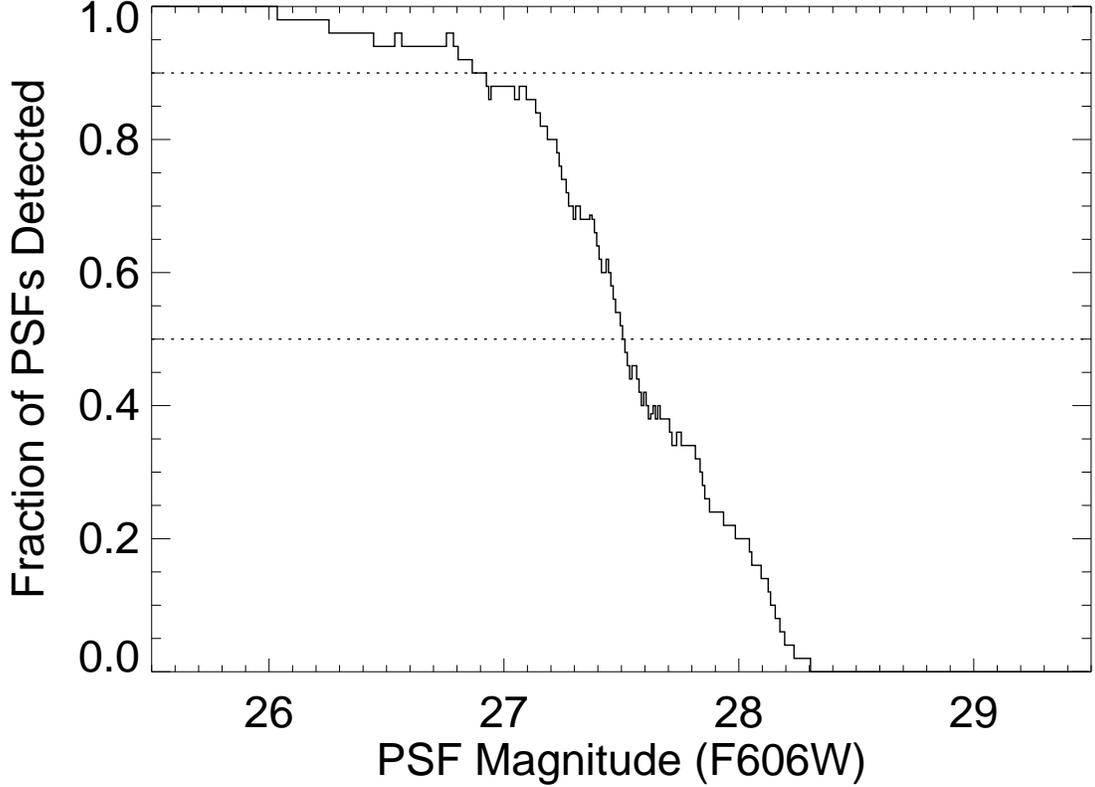}
  \caption[]{Detection efficiency as a function of STMAG magnitude for 
    PSFs more than 5\arcsec\ from Pluto. Points on the graph represent
    the running average of the detection efficiency, in bins 0.5 mag
    in width centered on points spaced every 0.01 mag. A total of 400
    synthetic PSFs, distributed randomly in the plane of the sky and
    uniformly in flux between magnitudes 25.5 and 29.5, were added to
    each of the four ACS/WFC images per \hst visit. Data from each
    visit were analyzed twice. The 90\% and 50\% level of detection
    efficiency are marked by horizontal dotted lines. The
    90\%-confidence magnitude limit is $m_{lim,90\%}=26.9$, while the
    50\%-confidence magnitude limit, i.e., the limiting magnitude as
    defined by \cite{Harris90}, is $m_{lim,50\%}=27.5$. These
    magnitude limits use the STMAG magnitude system with the F606W
    passband. Magnitude limits transformed into the Johnson V passband
    are presented in Table~\ref{sat_table}.
    \label{detection_fig}}
\end{figure}

The efficiency of detecting the synthetic point sources planted in the
ACS images is used to estimate the sensitivity of our search. The
detection efficiency, as a function of PSF magnitude (in the F606W
passband), is shown in Fig.~\ref{detection_fig}. Defining the limiting
magnitude to be the level at which the detection efficiency drops to
90\%, we find that the limiting magnitude, in the ACS/WFC F606W
passband and using the STMAG magnitude system \citep{Koorneefetal86},
of our search is $m_{_{F606W}}=27.5$. If we adopt a less stringent
definition of limiting magnitude as the magnitude where the detection
efficiency drops below 50\% \citep{Harris90}, then the limiting
magnitude of our search is $m_{_{F606W}}=26.9$.
\placefigure{detection_fig}

Both Pluto and Charon are severely over-exposed in the 475-second
integrations. Scattered light from these objects significantly
degrades the sensitivity of the ACS/WFC satellite search within
5\arcsec\ ($1.1 \times 10^5$~km) of Pluto. Since the plate scale of
the ACS High Resolution Channel (HRC) is roughly twice that of the
ACS/WFC (the ACS/HRC platescale is $\sim$0\farcs025/pixel versus
$\sim$0\farcs049/pixel for the ACS/WFC), it is less severely affected
by scattered light from Pluto and Charon, and so the February 2006
ACS/HRC observations were used to search for potential satellites
within 5\arcsec\ Pluto. Between 1\arcsec--3\arcsec\ (a projected
distance of $2.3 \times 10^4$--$6.9 \times 10^4$~km) the
90\%-confidence limiting magnitude is $m_{_{F606W}}=25.8$
($m_{_{F606W}}=27.0$ for 50\%-confidence), while between
3\arcsec--5\arcsec\ ($6.9 \times 10^4$--$1.1 \times 10^5$~km) from
Pluto the 90\%-confidence limiting magnitude is $m_{_{F606W}}=26.5$
($m_{_{F606W}}=27.3$ for 50\%-confidence). Between 0\farcs3--1\arcsec\ 
($6.9 \times 10^3$--$2.3 \times 10^4$~km) from Pluto the
90\%-confidence limiting magnitude is $m_{_{F606W}}=24$.

\subsection{Conversion of STMAG to V magnitudes}

The above magnitude limits use the STMAG magnitude system with the
ACS/WFC and HRC F606W filters \citep{Koorneefetal86}. These can be
converted into standard Johnson V magnitudes via the following
equation:

\begin{equation}
  \label{johnsonVeqn}
  m_{_V}=m_{_{F606W}} + c_{_0} + c_{_1} (\bv) + c_{_2} (\bv)^2 - Z_{_{ST}}
\end{equation}

\noindent where $m_{_{F606W}}$ is the magnitude in the F606W 
passband using the STMAG system and $\bv$ is the object's color in the
Johnson system. The coefficients $c_{_0}$, $c_{_1}$, and $c_{_2}$ as
well as the magnitude system zero point, $Z_{_{ST}}$ are given by
\citep{Siriannietal05}. Objects in the outer solar system exhibit a
wide range of \bv\ colors, e.g. $\bv=0.65$ for P1 and P2
\citep{Sternetal06IAUC} and $\bv=1.23$ for 5145 Pholus
\citep{Baruccietal05}. Since all three of Pluto's known satellites
exhibit roughly neutral colors (Charon has a \bv\ of 0.71
\citep{Buieetal97} compared with the solar \bv\ color of 0.67
\citep{Hardorp80}) it is reasonable to assume that any as yet
undetected satellites of Pluto would have $\bv \approx 0.7$.
Substituting the appropriate values into Eq.~\ref{johnsonVeqn}, we
find $m_{_V}-m_{_{F606W}}=-0.096$ for the ACS/WFC and
$m_{_V}-m_{_{F606W}}=-0.092$ for the ACS/HRC. If, instead, the
undetected satellites have extremely red \bv\ colors (i.e. similar to
5145 Pholus), then $m_{_V}$ would be $\approx 0.2$ mag fainter. The
limiting magnitudes of the satellite search, converted into Johnson V
magnitudes are given in Table~\ref{sat_table}. \placetable{sat_table}

\subsection{Limiting satellite diameter}

Once a limiting magnitude has been determined, the diameter, in
kilometers, of a spherical satellite, in the absence of significant
limb darkening, can be derived via the following equation
\citep{Russell16}:

\begin{equation}
  \label{radius_eqn}
  d=2.99 \times 10^{8} \, r \Delta p_{_V}^{-1/2} \, 
  10^{(m_{\sun}-m_{_V}+\beta \alpha)/5}
\end{equation}

\noindent where $r$ and $\Delta$ are the distances from Pluto to 
the Sun and Pluto to the Earth, respectively, in units of AU; $p_{_V}$
is the geometric visual albedo; $m_{\sun}=-26.75$ is the V magnitude
of the Sun at a distance of 1~AU \citep{Colinaetal96}; $\beta$ is the
phase law (in mag deg$^{-1}$); and $\alpha$ is the phase angle of the
object (i.e. the Sun-object-observer angle). We assume the phase law
for potential satellites identical to that for Charon, i.e.
$\beta=0.0866\pm0.0078$ mag deg$^{-1}$ \citep{Buieetal97}.

\begin{figure}
  \plotone{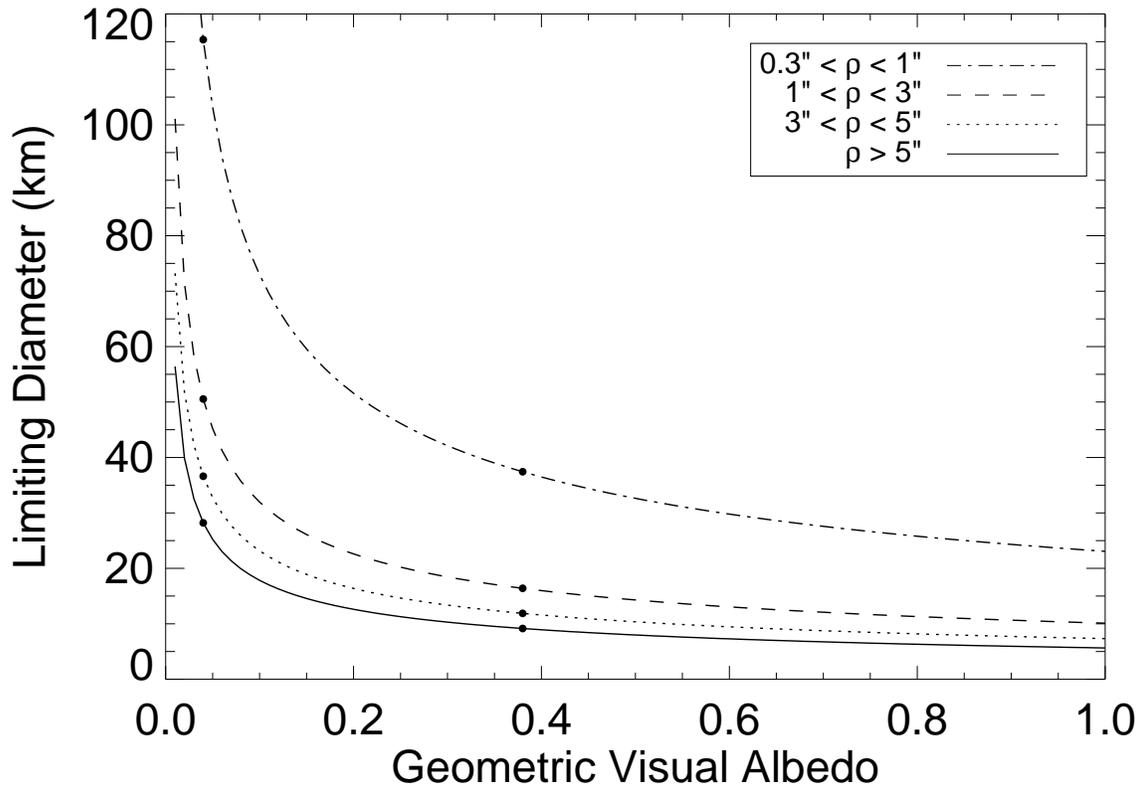}
  \caption[]{Maximum diameter of undiscovered satellites of Pluto, as 
    a function of geometric albedo, assuming a light curve amplitude
    of zero and a limiting magnitude defined by the 90\% detection
    efficiency criterion. Filled circles mark the locations of
    $p_{_V}=0.04$ (comet-like albedo) and $p_{_V}=0.38$ (Charon-like
    albedo).
    \label{radius_vs_albedo_fig}}
\end{figure}

Thus, assuming a very dark albedo of $p_{_V}=0.04$, comparable with
cometary nuclei \citep{Lamyetal04}, we can rule out, at the 90\%-level
of confidence, the existence of additional satellites in the Pluto
system larger than 49.4~km in diameter over the span of separations
from Pluto of 1\arcsec--3\arcsec, 36.1~km over the span of
3\arcsec--5\arcsec, and 28.6~km in diameter at separations of more
than 5\arcsec\ from Pluto. If, instead, we assume that potential
satellites are as reflective as Charon, i.e. having $p_{_V}=0.38$
\citep{Buieetal97}, then we can rule out satellites in these three
regions larger than 16.0, 11.7, and 9.3~km in diameter, respectively.
Within 1\arcsec of Pluto, the limiting diameters are 115~km for an
albedo of $p_{_V}=0.04$ and 37~km for an albedo of $p_{_V}=0.38$. Our
limiting diameters in this region are comparable to the limits
obtained by \cite{Sternetal94} using dynamical arguments and an
assumed orbital eccentricity of Charon of $10^{-4}$, which is
reasonable given the uncertainty of 7x10$^{-5}$ in the recently
published finding of zero eccentricity in the orbit of Charon
\citep{Buieetal06}. These results are summarized in
Table~\ref{sat_table}, and the limiting diameter for Plutonian
satellites for the four regions, as a function of satellite albedo, is
shown in Fig.~\ref{radius_vs_albedo_fig}.  \placetable{sat_table}
\placefigure{radius_vs_albedo_fig}

The above discussion has assumed a zero-amplitude light curve for
potential satellites. While Charon exhibits a relatively small light
curve amplitude (defined as the difference between maximum and minimum
magnitude and not the absolute deviation from the mean) of only 0.08
mag in V \citep{Buieetal97}, other Kuiper belt objects (KBOs) exhibit
much larger light curve effects \citep{Trilling:bernstein06}. An
extreme example is the KBO 2001~QG$_{298}$, which exhibits a light
curve with an amplitude of 1.14 mag in R \citep{Sheppard:jewitt04}.
If an object with a similarly extreme light curve exists within the
Pluto system and was at the minimum of its light curve during both of
the ACS/WFC (if the object is located more than 5\arcsec\ from Pluto)
or ACS/HRC visits (if the object is within 5\arcsec\ of Pluto), it
could have escaped detection, though its peak brightness would be
nearly 3 times greater than the upper limits quoted above. In this
pathological case, the length of the satellite in two dimensions could
be as large as the limiting diameters quoted above, while the length
in the third dimension could be up to a factor of 3 greater. The
effective diameter of such a cigar-shaped satellite would be
approximately 44\% greater than the above size limits.

The 90\%-confidence limit of $m_{_V}=25.7$ at separations greater than
1\arcsec\ from Pluto places an upper limit of roughly 49~km on the
diameter of undetected satellites in this region. Assuming bulk
properties (albedo, light curve, phase effect, density, etc) similar
to Pluto's smallest known moon, P2, potential undetected satellites in
this region must be less than 40\% the size of P2 with a mass of 2.5\%
that of P2. Such a small satellite would be unable to strongly perturb
the orbits of either P1 or P2, and therefore the proposed circular, or
near-circular, orbits of P1 and P2 \citep{Weaveretal06, Buieetal06} do
not necessarily preclude the existence of other very small satellites
in the Pluto system. Finally, we note that both P1 and P2 appear to be
in, or near, mean-motion resonance with Charon, and therefore, we
suggest that satellites below the detection limit of our search, may
occupy the other mean-motion resonances. We suggest further
observations with greater sensitivity to investigate this possibility.

\acknowledgments 

Financial support for this work was provided by the New Horizons
Pluto-Kuiper Belt mission. Additional support was provided by NASA
through grant numbers \mbox{GO-10427} and \mbox{GO-10774} from the
Space Telescope Science Institute, which is operated by the
Association of Universities for Research in Astronomy, Inc., under
NASA contract \mbox{NAS5-26555}.


\clearpage

\begin{deluxetable}{lccccc}
  \tablewidth{0pt}
  \tabletypesize{\footnotesize}
  \tablecaption{Observational Parameters \label{obs_table}}
  \tablehead{\colhead{Observation Date (UT)} & \colhead{Channel} & \colhead{Filter} & 
    \colhead{r (AU)} & \colhead{$\Delta$ (AU)} & \colhead{Phase Angle ($\alpha$)}}
  \startdata
  2005 May 15.045 & ACS/WFC & F606W           & 30.95 & 30.07 & 0.96$^{\circ}$ \\
  2005 May 18.141 & ACS/WFC & F606W           & 30.95 & 30.05 & 0.88$^{\circ}$ \\
  2006 Feb 15.659 & ACS/HRC & F606W           & 31.07 & 31.54 & 1.59$^{\circ}$ \\
  2006 Mar  2.747 & ACS/HRC & F475W and F606W & 31.08 & 31.31 & 1.77$^{\circ}$ \\
  \enddata
\end{deluxetable}

\begin{deluxetable}{lcccr}
  \tablewidth{0pt}
  \tabletypesize{\footnotesize}
  \tablecaption{Limits on Additional Satellites \label{sat_table}}
  \tablehead{\colhead{} & \multicolumn{4}{c}{Angular Separation from Pluto} \\
    \colhead{} & \colhead{0\farcs3--1\arcsec} & \colhead{1\arcsec--3\arcsec} & 
    \colhead{3\arcsec--5\arcsec} & \colhead{$>$5\arcsec}}
  \startdata
  Projected Distance (km) & $6.9 \times 10^3$--$2.3 \times 10^4$ & 
    $2.3 \times 10^4$--$6.9 \times 10^4$ & $6.9 \times 10^4$--$1.1 \times 10^5$ & 
    $> 1.1 \times 10^5$ \\
  50\%-Conf. Lim. V Mag &     & 26.9 & 27.2 & 27.4 \\ 
  90\%-Conf. Lim. V Mag & 24. & 25.7 & 26.4 & 26.8 \\ 
  Max. Diameter (km)\tablenotemark{a} & 37 & 16.0 & 11.7 & 9.3 \\
  $\rho_{_V}=0.38$ & & & \\
  Max. Diameter (km)\tablenotemark{a} & 115 & 49.4 & 36.1 & 28.6 \\
  $\rho_{_V}=0.04$ & 
  \enddata
  \tablenotetext{a}{Maximum satellite diameters calculated using
    Eq.~\ref{radius_eqn} and 90\%-confidence limiting magnitudes,
    assuming spherical satellites and no limb-darkening}
\end{deluxetable}

\end{document}